\documentclass{WileyMSP-template}

\usepackage{amsmath}
\usepackage[pagewise]{lineno}

\usepackage{cite}
\usepackage{setspace}

\begin{document}

\pagestyle{fancy}
\rhead{\includegraphics[width=2.5cm]{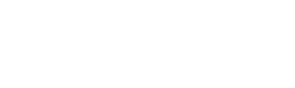}}

\title{Computational Design of Microarchitected Flow-Through \\ Electrodes for Energy Storage}

\maketitle


\author{Victor A. Beck*},
\author{Jonathan J. Wong},
\author{Charles F. Jekel},
\author{Daniel A. Tortorelli}, \\
\indent \author{Sarah E. Baker},
\author{Eric B. Duoss}, \emph{and}
\author{Marcus A. Worsley}


\dedication{}

\begin{affiliations}
V. A. Beck, J. J. Wong, C. F. Jekel, S. E. Baker, E. B. Duoss, M. A. Worsley\\
Lawrence Livermore National Laboratory,  Livermore, CA 94550\\
*Email Address: beck33@llnl.gov \\ 
\hfill \\
D. A. Tortorelli \\
University of Illinois, Urbana, IL 61801\\
Lawrence Livermore National Laboratory,  Livermore, CA 94550\\
\end{affiliations}



\begin{abstract}

\noindent Porous flow-through electrodes are used as the core reactive component across electrochemical technologies. Controlling the fluid flow, species transport, and reactive environment is critical to attaining high performance. However, conventional electrode materials like felts and papers provide few opportunities for precise engineering of the electrode and its microstructure. To address these limitations, architected electrodes composed of unit cells with spatially varying geometry determined via computational optimization are proposed. Resolved simulation is employed to develop a homogenized description of the constituent unit cells. These effective properties serve as inputs to a continuum model for the electrode when used in the negative half cell of a vanadium redox flow battery. Porosity distributions minimizing power loss are then determined via computational design optimization to generate architected porosity electrodes. The architected electrodes are compared to bulk, uniform porosity electrodes and found to lead to increased power efficiency across operating flow rates and currents. The design methodology is further used to generate a scaled-up electrode with comparable power efficiency to the bench-scale systems. The variable porosity architecture and computational design methodology presented here thus offers a novel pathway for automatically generating spatially engineered electrode structures with improved power performance.

\end{abstract}







\section{Introduction}

Electricity generated from renewable sources is a continually growing component of global energy production and a key driver for a sustainable energy future \cite{Chu:2012gd, Soloveichik:2015je, Chu:2017co}. Further expansion requires efficient and cost-effective integration into existing power distribution systems but intermittency and curtailment remain a challenge \cite{ Weber:2011eg, Li:2011fv, Arenas:2019bfa}. A number of strategies have emerged to address these issues, including electrochemical energy storage \cite{Gur:2018ct} and repurposing otherwise wasted electricity to electrify chemical manufacturing \cite{Schiffer:2017do, Kim:2019er, Shatskiy:2019jk, Noel:2019kh}. The direct electrochemical conversion of CO$_2$ is an especially powerful avenue as it simultaneously combines storage, chemical synthesis, and carbon-removal \cite{Ager:2018ce, Dinh:2018bk, Vedharathinam:2019ey}. As these advances are translated from the laboratory to industrial scale, energy efficient operation will become increasingly important to ensure economic viability \cite{Soloveichik:2015je,Gur:2018ct, Perry:2015ei,  Arenas:2019bfa}.

Porous flow-through electrodes are routinely used as the core reactor component across these applications and are ubiquitously employed for electrochemical energy storage using redox flow batteries \cite{Weber:2011eg, Soloveichik:2015je}. Flow battery performance is closely tied to the porous electrode properties. The electrode is often a disordered, homogeneous collection of micron-scale, electroactive particles like carbon fibers, felts or spherical substrates, any of which may be further coated with catalyst \cite{Weber:2011eg, Park:2016bt, FornerCuenca:2019fd}. These materials seek to maximize the surface reaction while minimizing overpotential and hydraulic losses. However, the material properties to meet these requirements are inherently adversarial and present a major challenge in attaining high performance. Open structures are necessary to allow fluid penetration, promote mass transfer, reduce pumping losses and supply reactants to the surface, but permeable geometries will reduce the solid fraction and require low hydrodynamicaly accessible surface area. In turn, the increasing porosity, decreasing  intrinsic surface area, and lower overall conductance lead to greater kinetic and Ohmic losses \cite{Weber:2011eg, FornerCuenca:2019fd}.  

Previous demonstrations of high power flow batteries have circumvented these issues by engineering the assembly to enable the use of very thin electrodes \cite{Aaron:2012er, Liu:2012dq}. The improved performance is attributed in part to the significantly decreased area specific resistances relative to thicker electrodes, like uncompressed carbon felts \cite{Brown:2016fj, Davies:2018ey}. Controlling electrode thickness and compression becomes an effective, bulk parameter to control the gross electrode microstructure, impacting average conductance, permeability, and mass transfer \cite{Brown:2016fj, Davies:2018ey, Banerjee:2019ei, Wong:2020bz}. 

To further drive performance, these architectures use sophisticated flowfields to appropriately distribute reactants across the electrode surface \cite{Darling:2014ks, Latha:2014ks, Houser:2017ko}. This important approach partially externalizes the challenges of balancing mass transport and electrochemical losses from the electrode to the fluid distribution system, providing further design freedom but at the cost of increased complexity. Previous studies have thus employed a combination of numerical \cite{Xu:2013bj, Knudsen:2015go, Ke:2018cs} and combined numerical and experimental \cite{Gerhardt:2018kg} approaches to develop engineering guidelines for flowfield channel dimensions and layouts to maximize peak power and efficiency. More recent work has employed X-ray computed tomography to simultaneously assess the impact of non-uniform compression and flowfield arrangement, thus connecting bulk engineering parameters to the electrode microstructure and its effective properties \cite{Kulkarni:2019kv}. Indeed, a growing body of research has focused on further establishing the connection between microstructure and hydraulic, mass transport, and electrochemical properties \cite{Wong:2020bz, Kulkarni:2019kv, Kok:2019jv,  FornerCuenca:2019br}.

As a complement to developing new assembly architectures, engineering the electrode structure directly is emerging as a viable route for improving performance \cite{Perry:2015ei, FornerCuenca:2019fd}. Holes made using laser perforation were used to create mass transport channels in carbon paper electrodes and increase peak power \cite{Mayrhuber:2014dc}. Similarly, slots milled into a large scale carbon felt electrode improved fluid distribution and decreased pumping losses without employing a costly flowfield \cite{Bhattarai:2019jr}. Dual-scale electrodes created by etching carbon papers \cite{Zhou:2016gq} or combining electrospun fiber mats with a backing layer \cite{Wu:2019dj} have enabled even more granular engineering of the structure. Similar dual-scale concepts have been introduced for lithium-ion electrodes \cite{Cobb:2014ba, Nemani:2015ka} and extended to create continuously variable porosity electrodes \cite{Golmon:2014je} which have been recently demonstrated to lead to improved rate capability while maximizing energy density \cite{Lu:2020hz}. However, to date, this novel idea has not been applied to make flow-through electrodes.

Additive and advanced manufacturing techniques can be employed to further extend and control the structural complexity of electrode materials \cite{Zhu:2016el, Lolsberg:2017jg, Arenas:2017gf, Hereijgers:2018hq}. Porous electrodes with superior mass transport have been created from carbon and graphene aerogels using direct ink writing for use in supercapacitors \cite{Zhu:2016el}. Flow through electrodes made from metals \cite{Lolsberg:2017jg, Hereijgers:2018hq}, including nickel and stainless steel \cite{ Arenas:2017gf}, have been produced at varying scales with complex, flow-controlling architectures. The resolution of the 3D printed, flow-through electrodes leads to feature sizes that exceed those of conventional electrodes \cite{FornerCuenca:2019fd} by 1-2 orders of magnitude.  However, a number of advanced manufacturing technologies exist with resolution approaching micrometers \cite{moran2016large, Hensleigh:2018fs} and tenths of micrometers \cite{Saha:2019fo} using materials that are, or can be readily transformed to be, suitable for use as electrodes.

The near arbitrary controlled provided by these techniques cannot be fully exploited without advanced analysis and design tools to guide the electrode architecture. Simulation has been used extensively and effectively to develop a more detailed understanding of the transport and electrochemical processes in flow batteries \cite{Esan:2020jb, Chakrabarti:2020bi}. The computational efforts have additionally provided design guidance, identified key control variables, and provided useful heuristics highlighting the importance of flow uniformity when engineering the electrode assembly \cite{EscuderoGonzalez:2014ia, JyothiLatha:2014gk, Zheng:2016fg, Yuan:2020bf}. 

The methodology necessarily involves beginning with an initial system architecture, analyzing the system, and then improving it through human-driven iteration. This process can be laborious and, crucially, explores only a limited portion of the design phase-space. A novel, alternative technique is to use automatic design algorithms, like topology optimization\cite{Sigmund:2013ht}, to invert the design process and aide in the phase-space exploration. Instead of evaluating the performance of a proposed architecture, a performance target is specified and the algorithm iterates over permissible architectures to meet the target. This can lead to intriguing, non-intuitive designs which are nevertheless high performance, as have been recently demonstrated for flowfield design in flow batteries \cite{Chen:2019kx} and fuel cells \cite{Behrou:2019gx}. Similar inverse design concepts have also been used to optimize the porosity of lithium ion batteries \cite{Golmon:2014je} but have never been applied to design the structure of flow through electrodes. 

We introduce the concept of algorithmically designed, microarchitected variable porosity 3D flow through electrochemical reactors. As a specific application, we focus on energy storage by designing flow through electrodes for vanadium redox flow batteries. We begin by describing our modeling, simulation and optimization methodology, including using high resolution continuum simulation to develop a homogenized description of the constituent microstructure unit cell. High performance computing is then employed to determine optimal distributions of the spatially varying unit cell porosities to maximize power efficiency across operating conditions. The resultant architectures are evaluated for their power performance and compared to bulk, porous flow through electrodes. The mechanisms leading to improved power efficiency are identified and connected to the underlying electrode structure. Finally, we demonstrate how this computational design methodology can be used to scale-up electrodes while minimizing power efficiency losses. The design methodology provides a framework for automatically generating high performance, architected 3D electrodes which can fully exploit the design freedom from advanced and additive materials manufacturing techniques.

\section{Computational Design of Electrode Architecture }

\begin{figure}
\centering
\includegraphics[]{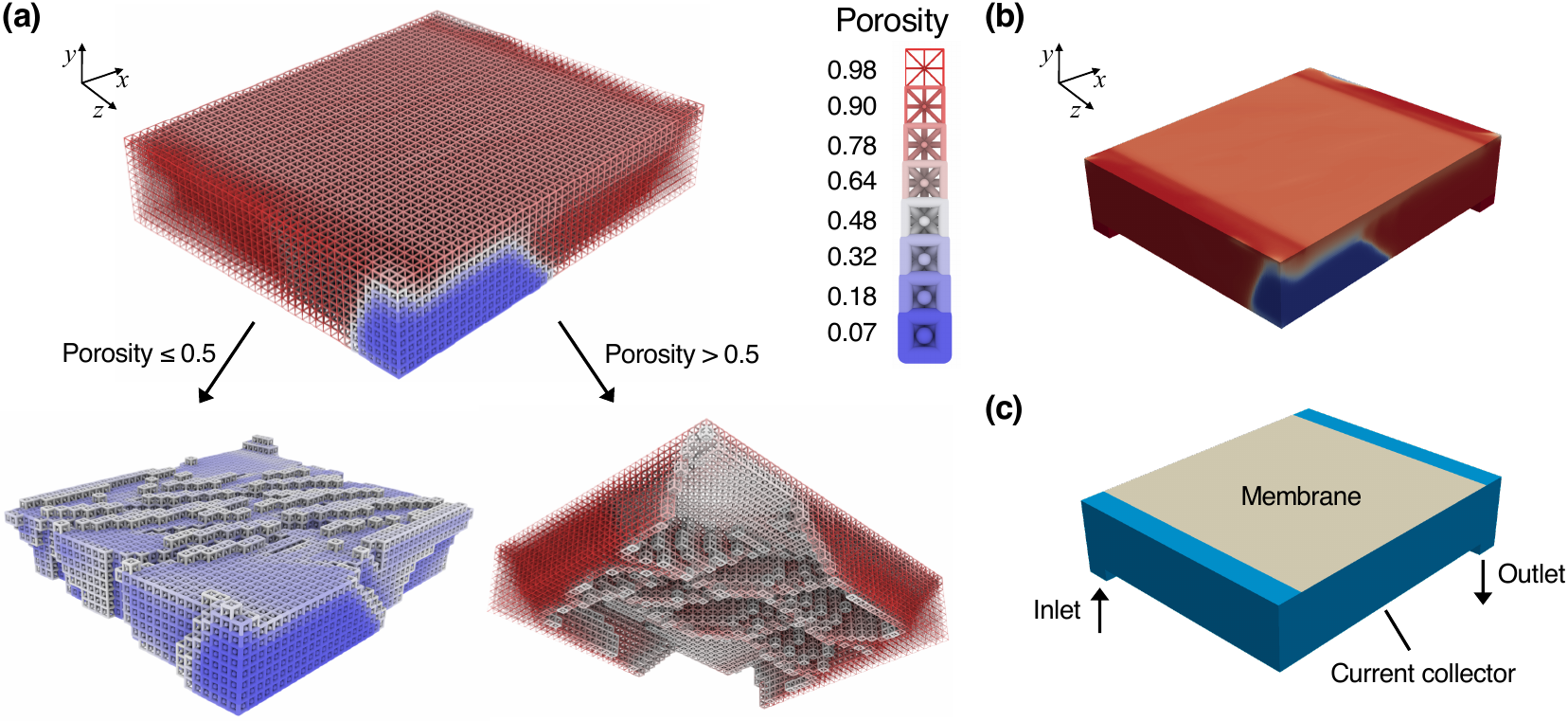}
\caption{(a) An example of a variable porosity, microarchitected electrode. The electrode is split into portions with unit cell with porosity above and below 0.5 for visualization. The porosity is controlled by fixing the unit cell length but allowing the rod diameter to vary.  The unit cell size is arbitrary and is chosen to be large in this example to aide with visualization. (b) The microarchitected electrode can be equivalently described with a spatially varying continuous porosity field (same legend as (a)). (c) The design domain is the negative half-cell compartment of a vanadium flow battery. The current collector and membrane have dimension of 2 cm x 2 cm. The inlet and outlet are square with area 0.02 cm x 0.02 cm. The compartment overall dimension is 2.4 cm x 2 cm x 0.5 cm.}
\label{fig:electrode}
\end{figure}

A combination of modeling, simulation and computational design optimization is used to generate the 3D, architected porosity electrodes presented in this work. The techniques described here are generally applicable to dilute, single-phase, flow-through porous electrochemical reactors. Here, this methodology is applied to determine optimal electrode architectures for electrochemical energy storage applications. Specifically, we model the electrode when used in the negative half-cell of a vanadium flow-through battery and seek to optimize the electrode architecture to minimize the power loss at fixed flow rates and discharge currents.

As shown in Figure \ref{fig:electrode}, the electrode is divided into regular, isotruss unit cells composed of rods/ fibers with equal radius. The fiber radius of these component cells is allowed to vary spatially. Previous advanced manufacturing work suggests scales and resolutions that would imminently permit the manufacture of porous carbon electrodes at a scale of 100 cm$^2$ with square unit cells of length $L=50$ $\mu$m \cite{Zhu:2016el, Hensleigh:2018fs, Saha:2019fo, moran2016large}. Assuming equal print resolution for both the void and the solid, the fiber radius, $r$, is arbitrarily permitted to vary from $r_{min} = 1.25$ $\mu$m to $r_{max} =11.25$ $\mu$m. The unit cell length is fixed, and thus the fiber radius can be used to directly control the porosity and surface area per volume of the unit cell, directly impacting the electrochemical, mass transport, and hydrodynamic response of the electrode as described below. The resultant designed lattice architecture can be readily transformed into a surface file and used as input for advanced manufacturing techniques. Note that though this work employ an isotruss geometry, these techniques are readily applicable to any unit cell structure.

\subsection*{Continuum simulation}
Following previous 3D simulations \cite{Ma:2011ee,Xu:2013bj, Wang:2014ky, Yuan:2020bf}, the negative half-cell of an all vanadium flow-through battery is modeled using porous electrode theory \cite{Newman:1975hn,trainham:1977da,Shah:2008ce}. Every point in the continuum represents both liquid and solid and is characterized by the local porosity as determined by the local rod radius of the unit cell, $\epsilon \equiv \epsilon\left(r\left(\vec{x}\right)\right)$. Note that because the unit cell rod radius changes with position, $r=r\left(\vec{x}\right)$, the transport properties of the system will also be position dependent. 

The electrolyte is a solution of $V^{2+}/V^{3+}$ in 1 M sulfuric acid. The mass balance expression for the reactive species $i\in\left\{V^{2+},V^{3+}\right\}$ is,
\begin{equation}
\vec{\nabla}\cdot\left( \vec{v}c_i-D_i\vec{\nabla}c_i\right)=aj_{n,i}, \label{eq:mb}
\end{equation}
where $c_i$ is the species concentration, $D_i$ is the effective diffusivity of the species in the liquid, $a$ is the specific area per volume, and $\vec{v}$ is the superficial velocity. Additionally, a high, constant conductivity solution is assumed and electromigration is ignored. The mass transfer flux from the solid is the product of the mass transfer coefficient and the difference between the surface and bulk concentrations: $j_{n,i} = k_m\left(c^s_i - c_i\right).$ The current density from the surface reaction ($V^{3+} +\:\: e^- \rightarrow \:\: V^{2+}, U_0$) is modeled using the Butler-Volmer expression,
\begin{equation}
i_n = k_0\left(c^s_{\text{V(II)}} e^{\beta\Delta\Phi} - c^s_{\text{V(III)}} e^{-\beta\Delta\Phi} \right), \label{eq:bv}
\end{equation}
where $k_0$ is the rate constant, $\beta\equiv F/2RT$, and we define $\Delta\Phi \equiv \Phi_1 - \Phi_2 -U_0$ using the solid, $\Phi_1$, and liquid, $\Phi_2$, potentials. The current density is related to the surface species flux through Faraday's Law, $i_n = F j_{n,V^{3+}} = -F j_{n,V^{2+}}$. The liquid and solid potentials are modeled using Ohm's law,
\begin{equation}
-\vec{\nabla}\cdot\left( -\sigma\vec{\nabla}\Phi_1\right)=\vec{\nabla}\cdot\left( -\kappa\vec{\nabla}\Phi_2\right)=ai_{n}, \label{eq:ohm}
\end{equation} 
where $\sigma$ and $\kappa$ are  the effective conductivities of the solid and liquid, respectively. Finally, the flow field in the porous medium is determined from the Navier-Stokes equation augmented with a Darcy drag,
\begin{equation}
\rho \vec{v}\cdot\vec{\nabla}\vec{v} + \frac{\mu}{\alpha}\vec{v} = -\vec{\nabla}p + \mu \nabla^2 \vec{v} \label{eq:flow} ,
\end{equation}
 where $\rho$ is the electrolyle density, $\mu$ is the electrolyle viscosity, $p$ is the pressure, and $\alpha$ is the position dependent permeability. Though it is included in the solution algorithm, the non-linear term makes a negligible contribution since the Reynolds number is small throughout the porous electrode, $Re \ll 1$.

The design domain in this work is the negative half-cell electrode compartment of a discharging vanadium flow through battery as show in Figure \ref{fig:electrode}c. For a specified flow rate, $Q$, and inlet area $A_{in}$, the velocity at the inlet is set to a uniform velocity normal to the boundary,
\begin{equation}
\vec{v}=-\frac{Q}{A_{in}}\vec{n}, 
\end{equation}
The inlet species concentration is fixed at $c_i =1$ M. The top portion of the domain is adjacent to the membrane, and a fixed current density is specified as the boundary condition,
 \begin{equation}
-\kappa\frac{\partial \Phi_2}{\partial \vec{n}} = \frac{I}{A},
 \end{equation}
where $I$ is the applied current and $A$ is the membrane area. The bottom boundary opposite the membrane is the surface of the current collector with boundary condition, $\Phi_1 = 0$. The domain exits to zero pressure, and all other flow boundary conditions are wall-type boundaries with no-slip boundary conditions for the velocity and no-flux boundary conditions for the pressure. All other scalar boundary conditions are no-flux (i.e., homogeneous Neumann) boundary conditions. 

\begin{table}[b]
\centering
  \caption{\ Physical Parameters}
  \label{tbl:parameters}
  \begin{tabular*}{0.48\textwidth}{@{\extracolsep{\fill}}llll}
    \hline
    Parameter & Value & Units & Ref \\
    \hline
    $D_{V^{2+},0}$ & 2.4 x $10^{-10}$ & m$^2$/s &\cite{Yamamura:2005ic} \\
    $D_{V^{3+},0}$ & 2.4 x $10^{-10}$ & m$^2$/s & \cite{Yamamura:2005ic}\\
    $\kappa_{0}$ & 40 & S/m & Estimate\\
    $\sigma_{0}$ & $10^4$ & S/m & \cite{LU:1993we}\\
    $U_{0}$ & -0.25 & V & \cite{Li:2011fv}\\
    $k_{0}$ & 1.75 x $10^{-7}$ & m/s & \cite{Shah:2008ce}\\
    $T$ & 300 & K & Assumed\\
    $\mu$ & 8.9 x $10^{-3}$ & Pa-s & Assumed\\
    \hline
  \end{tabular*}
\end{table}

\subsection*{Homogenization}

\begin{figure}
\centering
\includegraphics[]{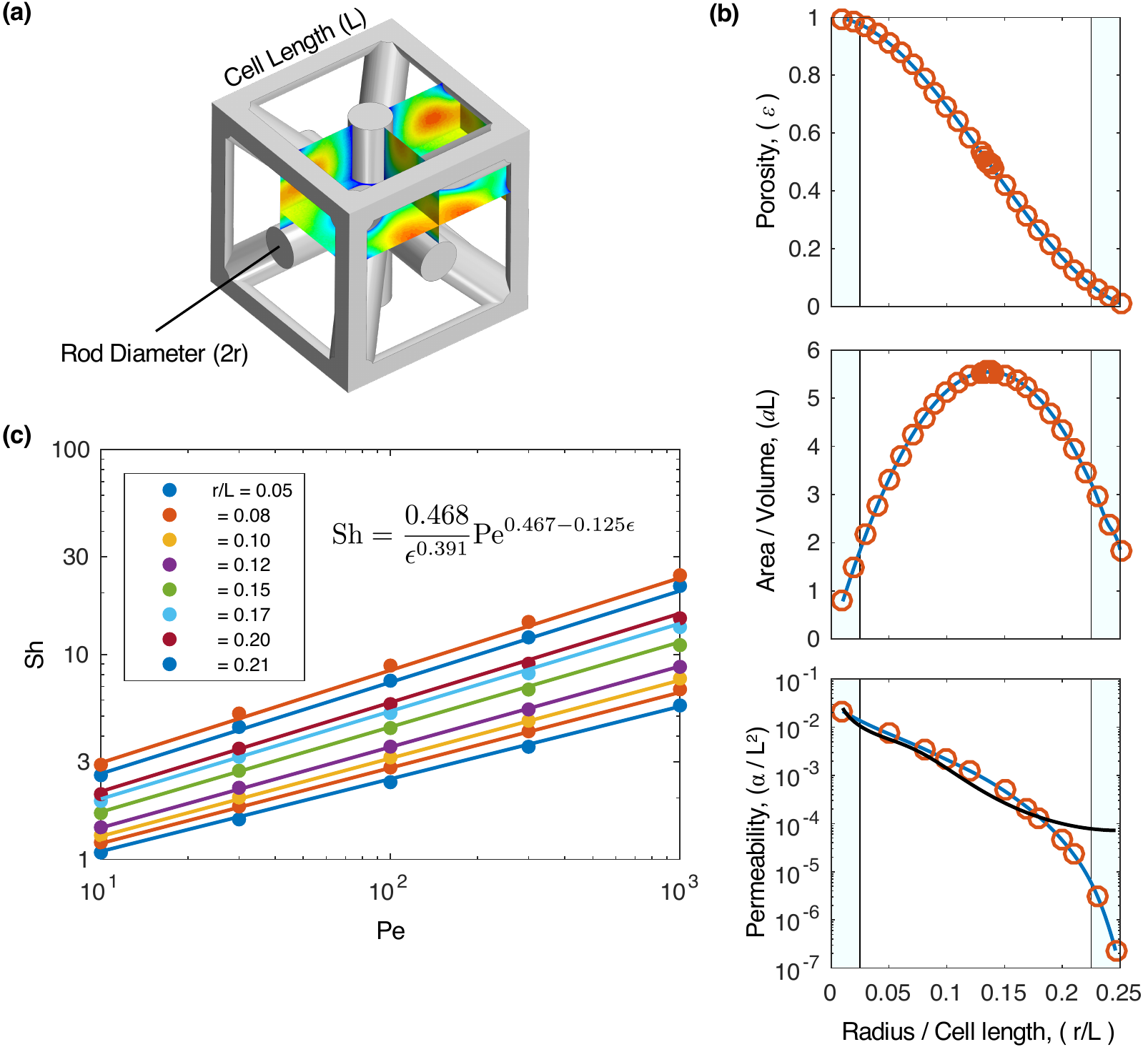}
\caption{(a) The isotruss unit cell is pictured with an illustrative planar slice of the resolved CFD computation in the interior of the cell. (b) The dependence of the porosity, area per volume, and permeability on $r/L$. The blue regions are outside of the range of permitted fiber radii given by $r_{min}$ and $r_{max}$ for the architectures designed in this work. The permeability plot includes a comparison, in black, to the correlation from Davies et al. for fibrous media \cite{davies:73bk}. The symbols are the results from CFD and the blue lines are spline fits (c) The unit cell Sherwood number, $Sh$, is presented as a function of the unit cell P\'eclet number, $Pe$, and the unit cell porosity, $\epsilon$. The symbols are the results from CFD and the curves correspond to the correlation equation, as shown.}
\label{fig:response}
\end{figure}

The electrochemical, transport, and hydrodynamic responses of the unit cells as a function of the rod radius must be determined to apply the porous electrode model. Briefly, a 3D resolved, microscopic model of the isotruss unit cell in Figure \ref{fig:electrode}a is developed using the commercial CFD software package Starccm+ (Siemens), a more detailed description of the procedure can be found in the Supporting Information and from our previous work \cite{Zhu:2018gr, Wicks:2021bn}. The intrinsic area per volume and porosity depend on the ratio of the internal rod radius to the fixed unit cell edge length, $L$. The CAD package in Starccm+ is used to calculate this relationship for several values of $r/L$. These points are then fit using a spline to produce the curves in Figure \ref{fig:response}b, and we emphasize that these properties are position dependent:
\begin{linenomath}
\begin{align}
a &=a \left(r\left(\vec{x}\right)\right), \\
\epsilon&=\epsilon\left(r\left(\vec{x}\right)\right).
\end{align}
\end{linenomath}
For the diffusive and conductive properties, the Bruggeman relation is employed\cite{Chung:2013gu,trainham:1977da,Shah:2008ce,Newman:1975hn}, and these properties are also position dependent, 
\begin{linenomath}
\begin{align}
D_i&=D_{i,0}\left[\epsilon\left(r\left(\vec{x}\right)\right)\right]^{3/2}, \\
\kappa&=\kappa_{0}\left[\epsilon\left(r\left(\vec{x}\right)\right)\right]^{3/2}, \\
\sigma&=\sigma_{0}\left[1-\epsilon\left(r\left(\vec{x}\right)\right)\right]^{3/2},
\end{align}
\end{linenomath}
with $D_{i,0}$ the molecular diffusivity of species $i$, and $\kappa_0$ and $\sigma_0$  are the conductivities of the liquid and solid, respectively. The physical parameter values used are listed in Table \ref{tbl:parameters}. 

The permeability is determined by meshing the void domain in the unit cell and calculating the steady, fully-developed velocity field for an applied pressure drop (recall $Re \ll 1$). Following Darcy's Law, the slope of the linear response of the superficial velocity, $\vec{v}$, against pressure is used to determine the permeability \cite{Tahir:2009gb}, 
\begin{equation}
\alpha=\alpha\left(r\left(\vec{x}\right)\right).
\end{equation}
Note that for the isotruss the permeability tensor is isotropic and characterized by a single scalar component. These values of permeability are also fit to a spline and are compared in Figure \ref{fig:response}b against correlations for fibrous porous media \cite{davies:73bk,Tahir:2009gb}. The correlations are expected to be accurate only for dilute fiber beds (i.e., high porosity), and there is good agreement with the calculation as the fiber radius to unit cell length ratio decreases.

The mass transport properties of the unit cell are determined by using the convection-diffusion equation to simulate the transport and surface-consumption of a dilute species assuming perfectly adsorbing fiber surfaces. Because creeping flow is assumed, there is no $Re$ dependence, and the only parameters are the species $Pe = r \left| \vec{v}\right| / D_{i,0}$ and the porosity. Simulations are performed across $Pe$ and $\epsilon$ to calculate the effective mass transfer coefficient,
\begin{equation}
k_m \equiv \frac{\left| \vec{v}\right|}{aL}\left(1-\frac{\left<c_{out}\right>}{\left<c_{in}\right>} \right),
\end{equation}
where $\left<\cdot\right>$ is the axial velocity average (i.e., mixing-cup average). The simulated values of the effective non-dimensional mass transfer coefficient, $Sh \equiv k_m r /D_0$, is then fit to yield the correlation shown in Figure \ref{fig:response}c. In the dilute limit, the expression reduces to $Sh \approx 0.468Pe^{0.342}/\epsilon $ and is in nearly exact agreement with a previous experimentally informed and frequently used correlation \cite{Geankoplis:2015jt,trainham:1977da,Darling:2014ks,Gerhardt:2018kg}. Note that because of the porosity and velocity dependence, the mass transfer coefficient is also position dependent:
\begin{equation}
k_m=k_m \left[  \epsilon \left(r\left(\vec{x}\right) \right),Pe \left(  \vec{v}\left(\vec{x}\right) \right)  \right].
\end{equation}

\subsection*{Optimization}

Using the continuum, forward model for the electrochemical and homogenized response of the system, the total power loss for any porosity distribution can be calculated. The power loss objective function is defined as the sum of the electric power losses, $P_{elec}$, and the hydraulic power losses, $P_{flow}$:
\begin{equation}
P_{tot} = P_{elec}+P_{flow}, \label{eq:obj}
\end{equation}
with,
\begin{equation}
P_{elec}=\int_{mem} \eta \;\vec{i}_2\cdot\vec{n}\; d\Gamma,
\end{equation}
an integral over the electrode-membrane interface, and
\begin{equation}
P_{flow}=\int_{in} \frac{p\;\vec{v}\cdot\vec{n}}{\Psi_{pump}} \; d\Gamma \label{eqPFlow},
\end{equation}
an integral over the inlet \cite{Gerhardt:2018kg}. The current density in the liquid is given by $\vec{i}_2$. The local overpotential is defined as, $\eta = \Phi_{1,cc} - \Phi_{2,mem} - U_0$, where $\Phi_{1,cc}\equiv0$ is the potential at the current collector, $\Phi_{2,mem}$ is the potential at the interface between the electrode and the membrane, and the Nernst potential only contributes the standard potential since the inlet concentrations of oxidant and reductant are set equal. The power efficiency is thus defined as $\Xi = 1 - P_{tot}/IU_0$.  In this work the pump efficiency is idealized and assumed to be $\Psi_{pump}=1$. Lower pump efficiency is equivalent to an increased weighting of the flow contribution to the objective function as seen in Equation (\ref{eqPFlow}). 

We seek to determine the distribution of unit cells with rod radius, $r\left(\vec{x}\right)$, which will minimize the total power loss, $P_{tot}$:
\begin{equation}
\begin{aligned}
&\min_{r\left(\vec{x}\right)} \:\: P_{tot}  \\
&s.t. ~~~ r_{min} \le r\left(\vec{x}\right) \le r_{max}.
\end{aligned}
\end{equation}
The total derivative, $dP_{tot}/dr\left(\vec{x}\right)$, subject to the constraints imposed by the physical model in Equation (\ref{eq:mb})-(\ref{eq:flow}) is calculated using standard techniques from PDE constrained optimization (see Supporting Information). Briefly, the continuous adjoint approach is employed to derive analytical expressions for the adjoint PDEs \cite{Othmer:2008ka}. This results in one adjoint PDE per forward model PDE. For a given solution of the forward model, the adjoint PDEs are numerically solved. The total derivitative (i.e., sensitivity) is then computed from the forward solution, adjoint solution, and partial derivative of the Lagrangian function with respect to the design variables. The porosity distribution is updated using the Method of Moving Asymptotes (MMA) \cite{Svandberg:1987xi}. Iteration continues until the average relative change in the objective function over the last 5 steps is less than $10^{-4}$, to arrive at a local optimum for the rod distribution. A Helmholtz filter is used to regularize the optimization problem and control the smoothness length scale of the porosity variation \cite{Lazarov:2010ew}. The length scale parameter is set to $200$ $\mu$m.

The forward simulation, adjoint calculation, and gradient descent update are all performed using bespoke code written using OpenFOAM. The domain is meshed using $\approx 1.2M$ cubic, finite-volume cells. 

\section{Results and Discussion}

The ultimate power efficiency of a flow through electrode is engineered by balancing the losses arising from insufficient mass transport to the reactive surfaces against the hydraulic power necessary to drive fluid to those surfaces and provide charge conduction pathways. Below we characterize the power losses in flow through electrodes composed of isotrusses and apply design optimization to a three-dimensional model of coupled fluid flow, species transport, and current distribution, to address this engineering challenge.

\subsection*{Characterizing power losses in flow through electrodes}

\begin{figure}
\centering
\includegraphics[]{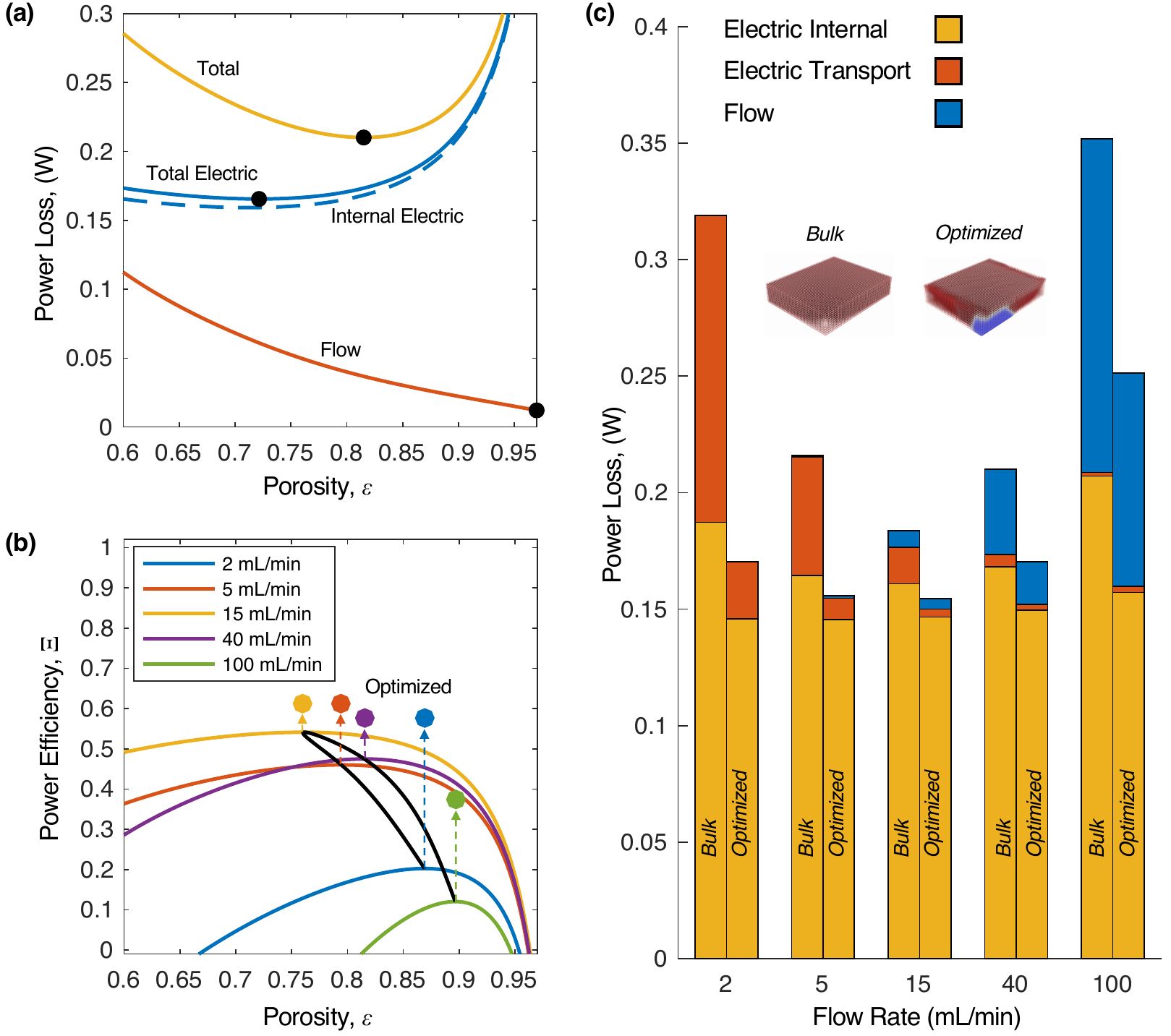}
\caption{(a) The total power loss  is separated into total electric and flow components for a bulk electrode, characterized by a single porosity, operated at 40 mL/min and 400 mA/cm$^2$. The total electric power loss is further separated into the portion due to concentration polarization and the portion due to internal electric resistance. The location of the minimum from each contribution is marked. (b) The power efficiency at a fixed current density of 400 mA/cm$^2$ as the porosity and flow rate are varied. The solid lines show the power efficiency for a bulk, monolithic electrode. The black curve is the loci of power efficiency maxima. The arrows show the increase in efficiency for the optimized electrode relative to the best performing bulk electrode. (c) The total power loss, separated into hydraulic losses, internal electric losses, and electric transport losses is presented for both the bulk and optimized electrodes at fixed current density of 400 mA/cm$^2$. The histogram couplets correspond on the left to the power losses for the bulk electrode at its optimal porosity (i.e., the maxima in (b)) and the right to the optimized, architected, variable porosity electrode.}
\label{fig:pl01}
\end{figure}

The porosity dependence of the total power loss in a bulk, monolothic electrode used as the negative half-cell of a vanadium flow-through battery (i.e, a homogeneous electrode which can be described by a single value for the porosity) is presented in Figure \ref{fig:pl01}a. At a fixed input flow rate of 40 mL/min and operating current density of 400 mA/cm\textsuperscript{2} the minimum power loss occurs when using an electrode with bulk porosity $\epsilon = 0.815$. Equivalently, the electrode is a uniform set of lattices with rod radius $r = 3.7$ $\mu$m. The total power loss is decomposed via Equation (\ref{eq:obj})-(\ref{eqPFlow}) into contributions from hydraulic losses and electric losses to reveal that this minimum results from a trade-off between each component. The flow losses decrease as the porosity of the electrode increases, with a minimum hydraulic power loss occurring at the highest attainable porosity. This is as expected, as the more open structure will have higher permeability. Alternatively, $\epsilon \rightarrow 1$ represents a singular limit for the electric losses since both Ohmic and kinetic overpotential losses grow unbounded as the solid material disappears. Instead, the electric losses generally decrease with increasing solid fraction since this increases reactive area and the effective solid conductivity, lowering Ohmic and surface overpotential losses. 

The electric losses pass through a minimum at  $\epsilon = 0.711$ and eventually begin to increase as the porosity further decreases. Increasing the solid fraction hinders transport to the reactive surface by displacing fluid, changing the local mass transfer coefficients, and decreasing the liquid conductivity. To better differentiate these effects, the internal electric loss, $P_{int}$, is determined by simulating an electrode wherein we prescribe the surface concentrations of all species to be equal to the inlet concentrations, reducing the problem to the solution of Equation (\ref{eq:ohm}). This idealization removes all concentration variations and hence all concentration polarization losses, and it is conceptually equivalent to canonical work on one-dimensional porous electrode models \cite{Newman:1962kt}. The concentration polarization power loss is thus defined as,
\begin{equation}
 P_{trans} \equiv P_{elec} - P_{int}.
 \end{equation}
 The internal electric loss has contributions only from Ohmic and kinetic overpotential losses \cite{Newman:1975hn,Newman:1962kt}, while the concentration losses will include contributions from variations in both the concentration \emph{and} the mass transfer coefficients. We thus equivalently refer to the concentration polarization losses as electric losses due to insufficient species transport, or simply electric transport losses. The internal power loss plotted in Figure \ref{fig:pl01}a is the dominant contribution to the total electric power loss and the dominant loss in general. At this flow rate and current, the transport losses are generally too small to impact the optimal porosity, but it is noted that these losses increase as the void volume is reduced.
 
 \subsection*{Minimizing power losses in flow through electrodes}
 
The total power efficiency, $\Xi$, in bulk electrodes operating at flow rates ranging from 2-100 mL/min is shown in \ref{fig:pl01}b. All of the power efficiency curves are similar and have a single optimal porosity that balances the flow and electric power losses. A spline fit to the loci of the curve maxima yields the expected optimal porosity as the flow rate is varied. At 400 mA/cm$^2$ the most efficient bulk electrode operates at the maximum of this spline curve. Equivalently, the minimum power loss is attained when operating at 15 mL/min with a bulk electrode porosity of $\epsilon = 0.761$.  

The key contribution in this work arises from allowing the geometry to vary spatially throughout the electrode by changing and optimizing the rod radius in each fixed unit cell. The optimization approach discussed above is used to determine a distribution of unit cell rod radii, $r\left(\vec{x}\right)$, which minimizes the total power loss. At each flow rate and current density, the initial input to the optimization process is the corresponding minimum bulk porosity. As shown in Figure \ref{fig:pl01}b, in each case the optimization algorithm is able to find a distribution of radii which leads to further decreases in the total power loss and hence greater efficiency than the bulk electrode. Indeed, the highest efficiency observed in Figure \ref{fig:pl01}b is $\Xi = 0.614$ for the optimized electrode operating at 15 mL/min, an efficiency increase over the bulk electrode of 13.5\%. Across flow rates, the efficiency relative to the best performing bulk electrodes is improved by as much as 310\%. This optimization procedure and analysis is repeated at current densities of 100 mA/cm\textsuperscript{2} and 200 mA/cm\textsuperscript{2} as presented in the Supporting Information, and, in general, lower current operation leads to even higher efficiency.

\begin{figure}
\centering
\includegraphics[]{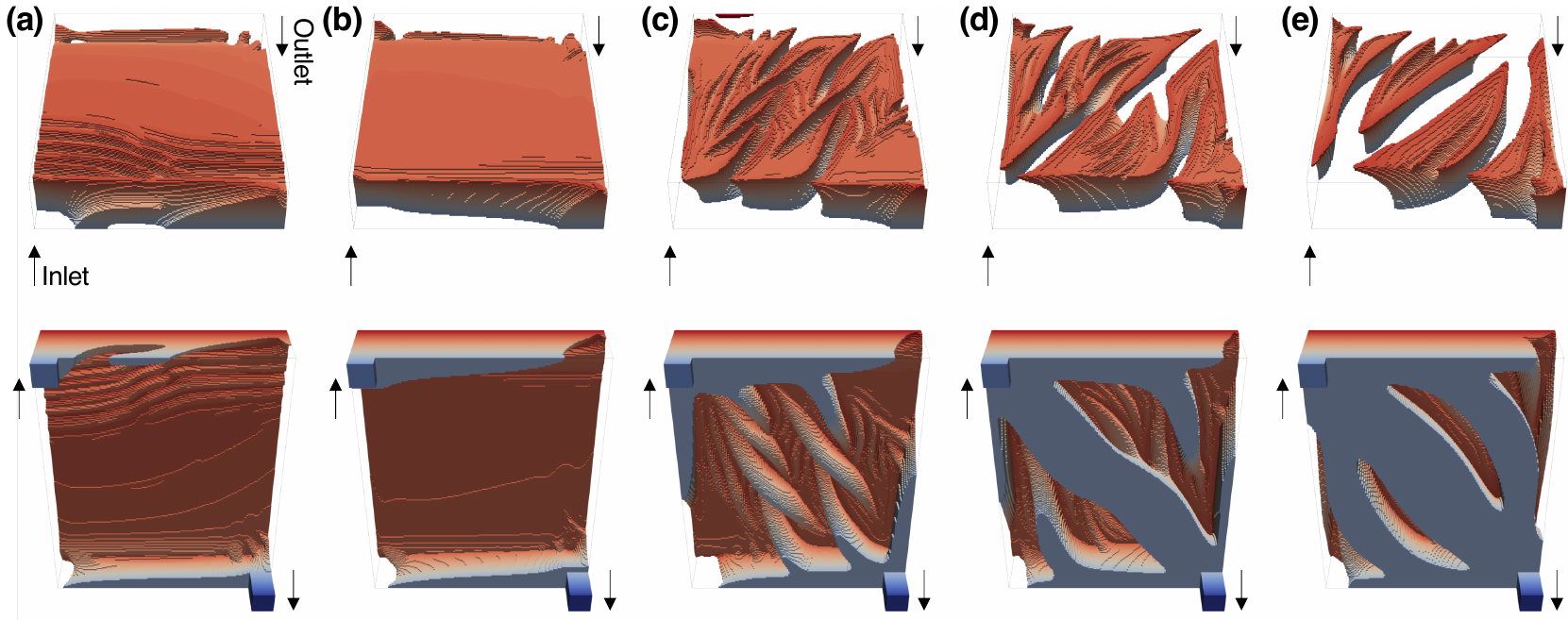}
\caption{The architected porosity electrodes resulting from the optimization procedure at a fixed current of 400 mA/cm\textsuperscript{2} and a flow rate of (a) 2 mL/min, (b) 5 mL/min, (c) 15 mL/min, (d) 40 mL/min, and (e) 100 mL/min. To aid in visualization, the electrodes are split into a ``solid'' half with $\epsilon$ $\le$ $0.5$, presented on the top row, and a ``void'' half where  $\epsilon$ $>$ $0.5$ presented on the bottom row. The electrodes are colored by height, with red adjacent to the membrane and blue adjacent to the current collector.}
\label{fig:geometry}
\end{figure}

The resultant locally optimal, variable porosity electrode geometries are visualized by splitting the electrode into two volumes. The ``solid'' portion of the electrode lumps all unit cells with $\epsilon \le 0.5$ while the ``void'' portion of the electrode lumps all unit cell with $\epsilon > 0.5$. The former is more closed to fluid flow and will behave more like a pure solid (i.e., preventing flow and enhancing electrical conduction), while the latter is more open and will behave more like a void (i.e., permitting flow and reducing electrical conduction). The two portions of the designed electrode are presented in Figure \ref{fig:geometry}, and it is emphasized that though we are visualizing solid blocks, all regions in the electrode are composed of unit cells (see Figure \ref{fig:electrode}a). 

\begin{figure}
\centering
\includegraphics[]{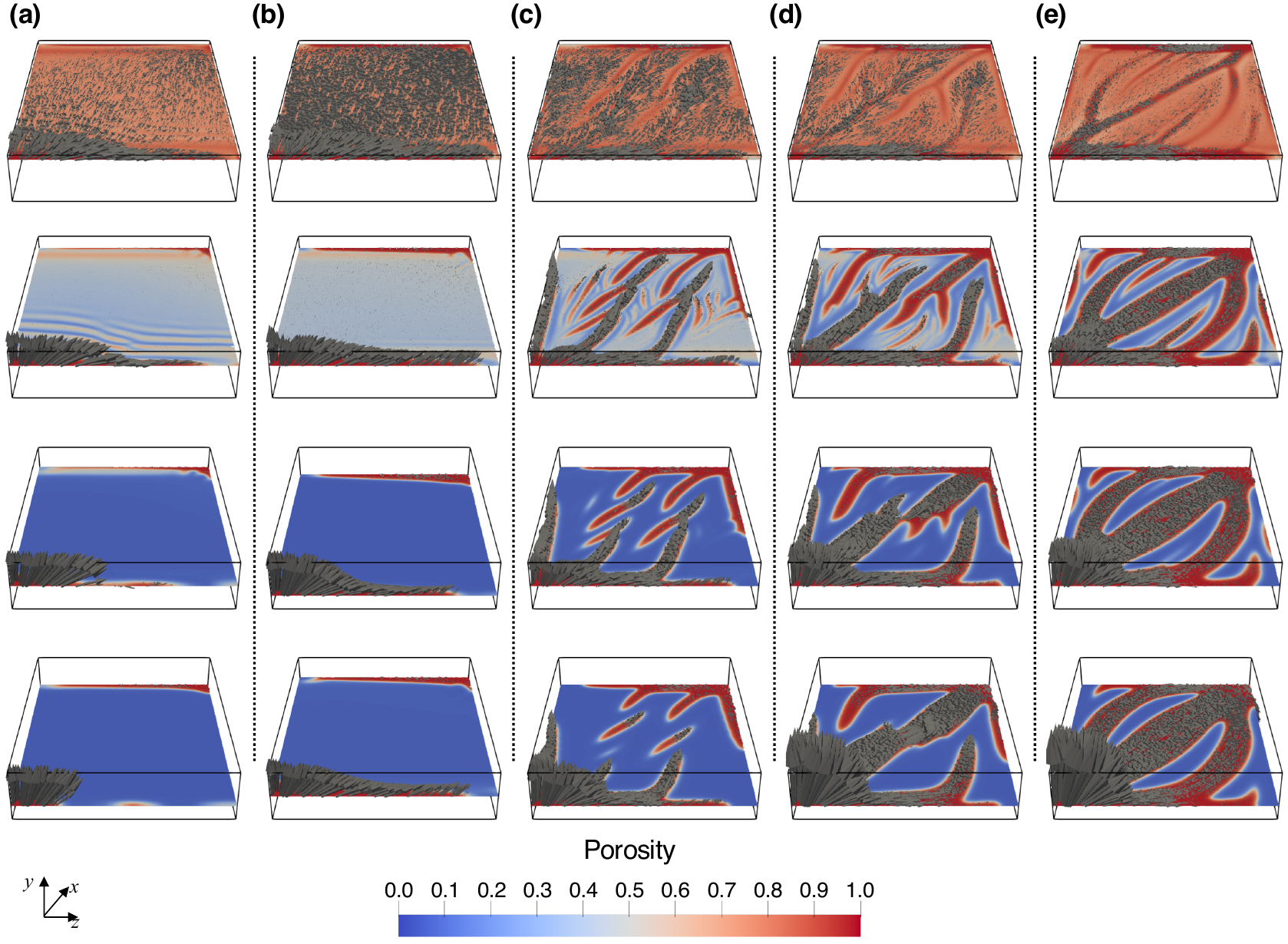}
\caption{Planar slices of the electrode porosity parallel to the current collector and membrane are presented for the architected electrode designed at flow rate of (a) 2 mL/min, (b) 5 mL/min, (c) 15 mL/min, (d) 40 mL/min, and (e) 100 mL/min. The planes are created at 1.5 mm, 2.5 mm, 3.5 mm, and 4.5 mm from the current collector. The velocity vector field is also presented, with vector length proportional to relative velocity magnitude within each column.}
\label{fig:slices}
\end{figure}

Representative slices of the porosity distribution and the flow vector field in planes parallel to the current collector are presented in Figure \ref{fig:slices}. At low flow rates, the optimal electrode geometry tends towards a large, mostly solid block with a thin high-flow region near the membrane. While at higher flow rates several large channels are carved into the solid block, and these grow larger with increasing flow rate. Across the designed electrodes, the porosity increases with distance from the current collector and all electrodes show a thin, high porosity, distributed flow region adjacent to the membrane.

The highest efficiency, optimized electrode is pictured in Figure \ref{fig:geometry}c and Figure \ref{fig:slices}c. It is a mostly solid structure near the current collector with several low porosity channels tunneling through the block that direct fluid from the current collector to the membrane. Near the membrane the number of channels increases, eventually leading to the high porosity region and appearing to spread the fluid along the membrane surface. In contrast, the electrode operated at 5 mL/min (see Figure \ref{fig:geometry}a and Figure \ref{fig:slices}a) begins by distributing the majority of the fluid along the top, upstream edge of the electrode. The channels in the highest effiency electrode thus appear to increase the number of injection points for the fluid and distribute them throughout the volume. In essence, the variable porosity electrode is performing an analogous function to a flowfield plate \cite{Darling:2014ks, Latha:2014ks, Houser:2017ko,Xu:2013bj, Knudsen:2015go, Ke:2018cs, Gerhardt:2018kg}, but now the flow control is integral to the electrode. Similar flow-directing structures are seen in Figure \ref{fig:geometry}d and Figure \ref{fig:slices}d, but, interestingly, these are most pronounced in the best performing, architected electrode.

 \subsection*{Comparing power losses in bulk electrodes to variable porosity electrodes}

The contributions to the total power loss for the bulk electrodes are presented in the first bar of each pair in Figure \ref{fig:pl01}c. At low flow rates, the electric transport losses are major contributors to the total loss in the bulk electrodes. There is insufficient mass transfer at low flow rates. Alternatively, at high flow rates the electric transport losses are small but at the cost of increased hydraulic losses. At each of these extrema the performance of the bulk electrode is maximized by using a more open porosity. At low flow rates, low porosity is necessary to enable greater mass transport, while at high flow rates it is instead the reduced flow resistance from higher permeability which is necessary to improve performance. Unfortunately, the higher porosity also leads to lower overall conductivity and an increase in internal electric losses. The loci of maxima in Figure \ref{fig:pl01}b can thus be understood as an optimization over operational flow rate to find the lowest porosity value which minimizes losses associated with fluid distribution: Operation at 15 mL/min \emph{enables} the use of the more solid, conductive electrode.

The architected electrodes are compared to the best performing bulk electrodes (i.e., the electrodes corresponding to the minima of the solid curves in Figure \ref{fig:pl01}b) operating at the same current and flow rate. In all cases, the variable porosity, optimized electrodes lead to a reduction in all contributions to the power loss. At the low flow rates the electric transport losses due to insufficient species flow are minimized by forcing all of the flow into a thin region adjacent to the membrane, effectively creating a thin, porous electrode (i.e., Figure \ref{fig:geometry}a). For high flow rates, an excess of flow amply supplies the reaction near the membrane but leads to large hydraulic losses which are minimized by carving by-passes in the thick bulk region (i.e., Figure \ref{fig:geometry}e). The internal resistance of the electrodes is also lowered by using the architected porosities. However, unlike for a bulk electrodes above, architected porosity relaxes the constraint between porosity and conductivity observed above for the bulk electrode allowing for increased power efficiency. Interestingly, the internal losses for the bulk electrode all appear to reach similar limiting values around 0.15 W. These are all lower than the internal losses observed in the bulk electrode and imply that variable porosity can improve performance even when mass transport losses are unimportant. Similar trends are observed at lower operating current densities (see Supporting Information).

\begin{figure}[tbhp]
\centering
\includegraphics{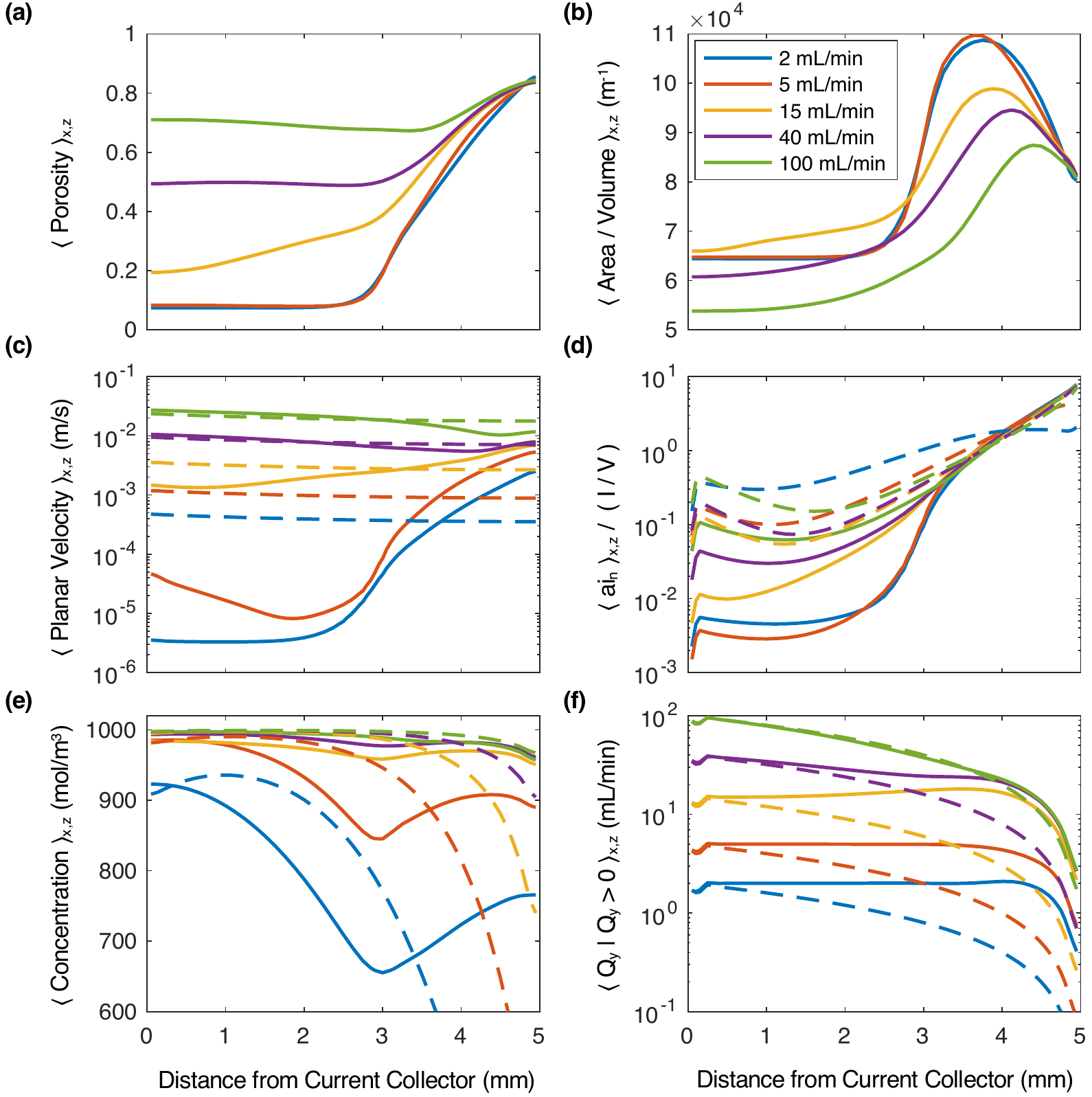}
\caption{Averages of the (a) porosity, (b) area per volume, (c) planar component of the velocity, $u_{x,z}$, (d) reaction current made dimensionless with the applied current divided by the electrode volume, $V$, and (e) the concentration of the reductant in planes parallel to the current collector and located at the given height above the current collector. In (f) the one-way flow rate defined in Equation (\ref{eq:Qpos}) is presented. The solid lines correspond to the variable porosity electrodes while the dashed lines correspond to the bulk electrodes at the optimal porosity (i.e., maxima of curves in Figure \ref{fig:pl01}b. The current density is fixed at 400 mA/cm\textsuperscript{2}.}
\label{fig:eluc}
\end{figure}

\subsection*{Characterizing the porosity distribution}

The porosity of the optimized electrodes is averaged on planes directly above the current collector as presented in Figure \ref{fig:eluc}a. At low flow rates, the planar averages show that the region near the current collector is near the lowest available value of the porosity, confirming that in Figure \ref{fig:geometry}a-b the ``solid'' regions are indeed monolithic and as solid as possible. The optimization algorithm has effectively created a conductive block extending from the current collector with a thickness of approximately 3 mm stacked below a 2 mm porous electrode with gradually increasing porosity. This is reminiscent of commonly recognized heuristics for engineering high power fuel cells and flow batteries which employ thin geometries with high porosity papers or felts \cite{Aaron:2012er, Liu:2012dq, Ke:2018dfa}. An important departure, however, is that in previous work the geometry is a thin flow channel with a constant porosity electrode. The variable porosity electrodes show that even for the thin, membrane adjacent 2 mm region, a smooth, gradual increase in porosity leads to greater power efficiency since the optimization procedure would adjust the porosity to be uniform near the electrode if this led to continued improvement. 

The gradual increase in porosity as the membrane is approached is observed across all of the optimized electrodes, and the transition occurs at approximately the same position, 3 mm. Further, the ultimate value of the porosity at the membrane is fairly constant across all architected electrodes, $\epsilon\approx0.84$. This value is within the range of optimal porosities determined for the bulk electrodes but exceeds the porosity of the best performing bulk electrode (e.g., the location of the maxima in Figure \ref{fig:pl01}b). The physical parameters of the system suggest that in the absence of concentration polarization a characteristic length scale for the bulk electrode is the penetration depth \cite{Newman:1975hn},
\begin{equation}
L_p=\sqrt{ \frac{RT}{ai_0F\left(\kappa^{-1}+\sigma^{-1}\right)}}
\end{equation}
which will vary with porosity, but for the range of porosities and intrinsic areas studied here is $\approx 0.1-0.9$mm. It is not obvious how to assign an equivalent characteristic length for the variable porosity electrodes, but the values are of the same magnitude as the characteristic length for the porosity variation in Figure \ref{fig:eluc}a. 

\subsection*{Elucidating mechanisms for enhanced performance}

The architected electrodes lead to higher power efficiencies across applied current densities and flow rates (see Supporting Information for additional results). The resultant flow, concentration and current fields for electrodes operating at 400 mA/cm\textsuperscript{2} are analyzed and compared to the bulk electrode fields to understand the factors that lead to these improvements. These quantities are again averaged on planes parallel and directly above the current collector and presented in Figure \ref{fig:eluc}.

The planar averages of the reaction current density, the magnitude of the righthand side of Equation \ref{eq:ohm}, normalized by the volume averaged current, $I/V$, are shown in Figure \ref{fig:eluc}d, revealing the spatial distribution of the electrochemical reaction and thus the electrode utilization. Driving this reaction at minimal power loss is the main goal of the electrochemical engineering problem posed in this work. Recall that for the bulk electrodes the highest power efficiency was attained for the bulk electrode operating at a flow rate of 15 mL/min.  When compared to the other bulk electrodes, it is evident from the  current distribution that in this electrode the reaction rate near the current collector is lowest and, because the operating current is fixed, the reaction rate near the membrane is greater. All of the architected electrodes exceed the power efficiency of this bulk electrode and additionally lead to lower internal losses. These electrodes all show further decreased reaction rate in the bottom portion (i.e., the region $<$3 mm from current collector), and consequently greater reaction rate near the membrane, relative to the bulk electrodes. High performance electrodes drive reaction closer to the membrane, and all of the optimized electrodes show exponential growth of the reaction rate as the membrane is approached. In the electrodes studied here, improved volumetric utilization of the electrode is not a requisite for improved power efficiency.

To maintain high reaction currents near the membrane, the reductant needs to be supplied at high molar fluxes, as quantified by the product of the local velocity and the concentration. The magnitude of the velocity projected onto the averaging plane is presented in Figure \ref{fig:eluc}b. For the bulk electrodes the nearly constant average value shows that the flow is approximately evenly distributed with distance from the current collector. All of the curves are similar and scale with the applied flow rate since the permeability is constant and nearly identical across the bulk electrodes and the flow equations are linear at low $Re$. Because at low $Re$ the flow vector field orientation does not change if the porosity is uniform, the only avenue for increasing reductant concentration near the membrane is to increase the flow rate. Indeed, in Figure \ref{fig:eluc}c it is only at the highest flow rates that a uniform, high concentration is attained for the bulk electrodes. At lower flow rates there is a strong depletion of the active species and thus a strong increase in the electric transport losses (see Figure \ref{fig:pl01}c). 

In contrast, the optimized electrodes succeed in attaining higher concentrations of the reductant through improved flow management without incurring unacceptably large hydraulic losses. At low flow rates (2-15 mL/min), the planar averages in Figure \ref{fig:eluc}c show that the variable porosity electrodes drive most of the fluid against the membrane and, with the exception of the electrode designed at 100 mL/min, exceed the flow rate of the bulk electrode operating at the same flow rate. The large velocities seen near the current collector at the highest flow rates are due to the large bypass channels seen in Figure \ref{fig:geometry}d-e. 

The impact of the flow distribution in the optimized electrodes on the concentration is readily evident in Figure \ref{fig:eluc}e. The concentrations near the membrane of these architected electrodes are higher than their bulk counterparts, with the exception of the optimization at 100 mL/min where the optimized electrodes has led to a decrease in the concentration near the membrane. Consequently, the electric transport losses seen in Figure Figure \ref{fig:pl01}c are reduced relative to the bulk electrode at all flow rates except 100 mL/min, where there is oversupply of species. 

For planes parallel to the current collector, the vertical transport induced by the electrode can be determined from the flow rate crossing the plane in only one-direction: 
\begin{equation}
\left<Q_y|Q_y>0\right>_{x,z}\equiv\frac{1}{2}\int_{x,z} \left|\vec{v}\cdot\vec{e}_y\right| + \vec{v}\cdot\vec{e}_y  \;dA, \label{eq:Qpos}
\end{equation} 
where $\vec{e}_y$ is a unit vector normal to the current collector and pointing in the direction of the membrane. It is equivalent to an integral of the vertical component of the vectors in Figure \ref{fig:slices} and provides a metric to quantify the vertical exchange of material. Indeed, when plotted in Figure \ref{fig:eluc}f the average one-way flow rate, $\left<Q_y|Q_y>0\right>$, shows that the optimized electrodes all lead to greater flow towards the membrane relative to the bulk electrode, especially near the membrane. In contrast, for the bulk electrode the only mechanism to increase vertical transport is to increase the flow rate, but this leads to the wasteful increase of hydraulic losses throughout the electrode. Interestingly, as seen from the 40 mL/min and 100 mL/min variable porosity electrode curves, the flow of material to the membrane is not maximized, instead it is optimized toward an apparent limiting value that is nearly equivalent to the vertical transport seen in the 15 mL/min variable porosity electrode. Any supply in excess leads to unnecessary hydraulic losses (see Figure  \ref{fig:pl01}c).

\begin{figure}[tbhp]
\centering
\includegraphics[]{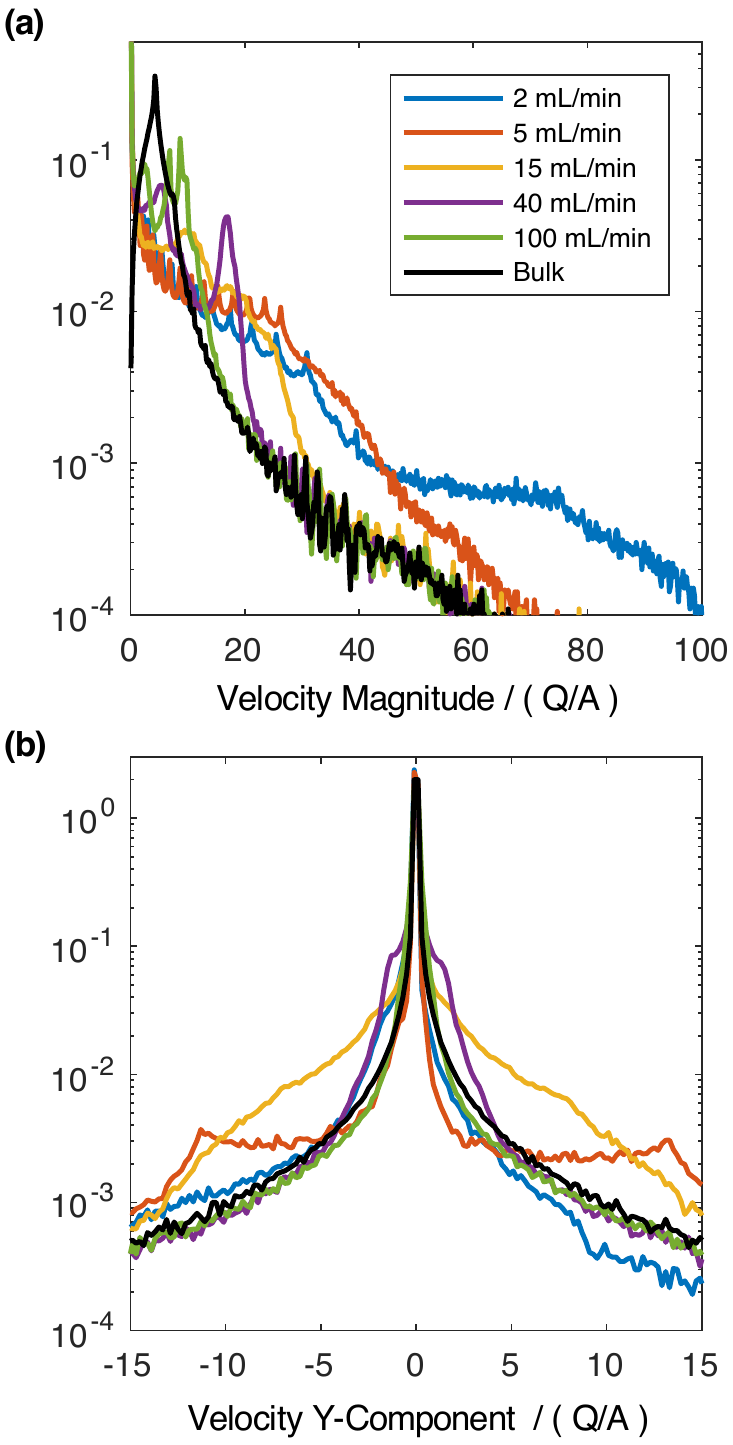}
\caption{The probability distribution function of (a) the magnitude of the velocity and (b) the perpendicular component of velocity, $v_y$,  as a function of the applied flow rate. The velocities are normalized by the area specific velocity, the ratio of the flow rate to the membrane area. The probability distribution functions for the optimal, bulk electrodes are presented in the black curves. Note that due to the linearity of the flow equations, the normalized flow field of the bulk electrodes is identical across flow rates and the probability distribution function for the bulk electrodes can be represented by a single curve labeled, ``Bulk.''}
\label{fig:velpdf}
\end{figure}

The central role of increased vertical transport is further supported by analyzing the probability distribution function of the vertical component of the velocity normalized by the area specific velocity, $Q/A$, as shown in Figure \ref{fig:velpdf}a. For all of the designed electrodes the velocity distribution is much broader, and this is especially pronounced for the electrode optimized at 15 mL/min. The 15mL/min optimized electrodes shows the most induced vertical flow. As shown in Figure \ref{fig:velpdf}b, the variable porosity electrodes lead to a broader distribution of high velocities relative to the bulk electrodes. In addition to enhancing convective transport, this serves to increase the local reaction rates by increasing the flow speed dependent mass transfer coefficients.

From these observations we can thus conjecture an interpretation of the resultant electrode architectures in Figure \ref{fig:geometry}. The optimization procedure is not only sculpting the gross features of the electrode, creating a thin, high porosity electrode near the membrane, it is also developing an integral flow distribution system, essentially a porous flowfield, to manage the flow paths. This is especially evident in the ``braided'' design seen in Figure \ref{fig:geometry}c, where we can hypothesize that the architectural features serve to distribute fluid vertically as well as to allow for maximal flow distribution, with minimal hydraulic loss, along the electrode. Of course, this is a single interpretation focused on flow distribution: The optimization procedure is general and algorithmic, and simultaneously optimizes flow paths, convective paths, and conduction paths to control reaction rates and automatically arrive at an optimal architecture irrespective of analogies to other engineering approaches.

\begin{figure}[t]
\centering
\includegraphics[]{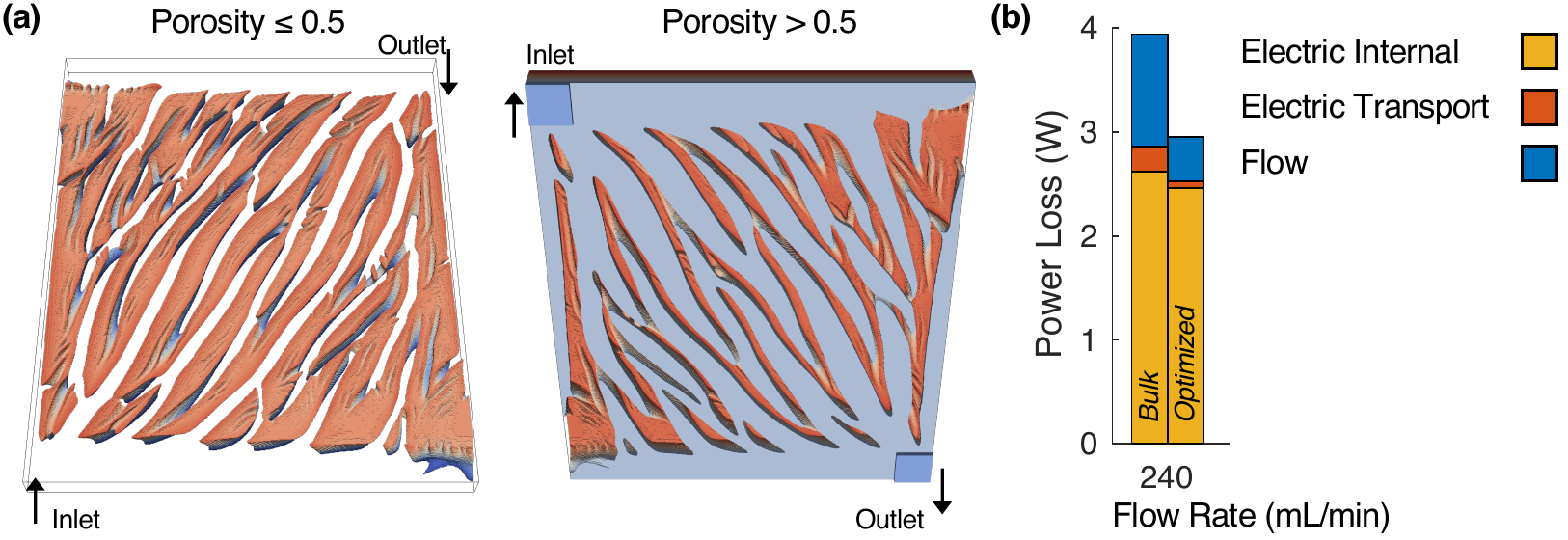}
\caption{(a) The architected porosity distribution for a scaled-up electrode designed for a half cell with dimensions 8 cm x 8 cm x 5 mm. (b) The total power loss, separated into hydraulic losses, internal electric losses, and electric transport losses for the best performing, scaled bulk electrode (first bar) and the optimized electrode (second bar).}
\label{fig:scaleup}
\end{figure}

\subsection*{Optimization of large scale electrodes}

The electrodes analyzed above are of typical experimental bench scale and have an electrode area of 4 cm$^2$. However, though these dimensions are expedient for analysis, it is often unclear how to translate performance insights realized at these smaller scales to the larger scales needed for industrial application. Industrially useful reactors are often several orders of magnitude larger \cite{Bhattarai:2019jr, Ke:2018dfa}. Thus, as a final demonstration of the utility of the approach described here, the planar dimensions of the half-cell compartment are scaled by 16x to a size of 8cm x 8cm x 5mm (i.e., electrode area of 64 cm$^2$) while keeping the current density and area specific velocity fixed. The mesh resolution is constant requiring the solution of an optimization problem with $\approx$20M design variables. The best performing bulk electrode at 4 cm$^2$ is scaled to this new dimension and compared to an architected electrode designed at this new dimension as shown in Figure \ref{fig:scaleup}a. 

When the bulk electrode is scaled in the absence of fluid distribution systems the power efficiency drops from $\Xi = 0.541$ to $\Xi = 0.384$, a 29.0\% reduction. Alternatively, the architected, optimized electrode experiences a much smaller reduction in power efficiency dropping from  $\Xi = 0.614$ to $\Xi = 0.539$, only a 12.3\% reduction and thus retaining nearly all of the power efficiency upon scale-up. Equally important, the scaled-up architected electrode exceeds the power efficiency of the scaled-up bulk electrode by 40.3\%. 

The enhanced performance is further reflected in the power losses presented in Figure \ref{fig:scaleup}b. The structures in the optimized electrode again lead to lower contributions to all loss components. A bulk electrode would, in practice, also use a fluid distribution system, but it is not possible for an external manifold to arbitrarily control the detailed local flow in the interior of the electrode. Indeed, even if it were possible the target flow distribution is generally unknown and would need to be determined via laborious and time inefficient testing and iteration. In this context, the advantages of the architected electrodes described here are self-evident, the variable porosity leads to expanded design flexibility that can be readily harnessed by the computational design algorithm to automatically generate structures with significantly enhanced performance.

\section{Conclusions}

This work has introduced the concept of algorithmically designed, microarchitected 3D flow through electrochemical reactors. The technique was used to design electrodes for the negative half-cell of a vanadium redox flow battery. Across flow rates, utilizing architected electrodes led to power efficiency gains of 13.5\% - 310\% over uniform porosity, bulk electrodes. Decomposing the power losses into flow losses, internal electric losses, and concentration polarization losses revealed that the algorithmically designed electrodes manipulate the porosity distribution to reduce the dominant loss contributors. These architectural changes included gross features like creating thin, high conductivity geometries at low flow rates, and creating large, low hydraulic resistance bypass channels for operation at higher flow rates. However, the best performing electrodes also included several more subtle architectural changes that shape both fluid and conductive pathways, creating an integrated fluid distribution system without sacrificing electrochemical performance. These conclusions were supported through analysis of the local reaction rates, flow, and concentrations, where it was demonstrated that the designed, variable porosity leads to improved transport from the current collector toward the membrane,  greater reaction adjacent to the membrane, and more vertical transport between the membrane and current collector. Algorithmically designed, architected porosity electrodes thus provide a novel pathway for electrode engineering without necessitating complex external manifolds like flowfields. As a final demonstration of the utility of this technique, the framework was employed to scale-up to 64 cm$^2$ from 4 cm$^2$. The scaled electrode incurred only a 12.3\% reduction in power efficiency and exceeded the power efficiency of the best performing bulk electrodes by 40.3\%. This work has thus provided a versatile, novel, variable porosity electrode architecture with vastly expanded design flexibility and a computational methodology to automatically generate high performance structures across length scales. The techniques described here are broadly applicable across flow-through electrodes and liquid-fed porous reactors.

\medskip
\noindent\textbf{Supporting Information} \par 
\noindent Supporting Information is available from the author.

\medskip
\noindent\textbf{Acknowledgements} \par 
\noindent This work was performed under the auspices of the U.S. Department of Energy by Lawrence Livermore National Laboratory under Contract DE-AC52-07-NA27344 and was supported by the LLNL-LDRD program under project numbers 16-ERD-051, 19-SI-005, and 19-ERD-035. LLNL Release Number LLNL-JRNL-819457.

\pagebreak

%

\bibliographystyle{MSP}
\bibliography{optimization}



\end{document}